\documentstyle {article}
\title{Noncommutative instantons: a new approach.} 
\author {Albert Schwarz \thanks{Partially  supported by  NSF grant 
No. DMS 9801009}\\
Department of Mathematics, University of California, \\
Davis, CA 95616 \\
schwarz@math.ucdavis.edu }    
\begin {document}
\maketitle
\begin {abstract}
We  discuss instantons on noncommutative four-dimensional Euclidean space.
In commutative case  one can consider instantons directly on Euclidean
space,
 then we should restrict ourselves to the gauge fields  that are gauge
equivalent to
the trivial field at infinity. However, technically it is more convenient
to work on four-dimensional sphere.
 We will show that the situation in noncommutative case is quite similar.
One can analyze instantons taking as a starting point the algebra
of smooth functions  vanishing at infinity, but it is  convenient to add a
unit
element to this algebra ( this corresponds  to  a transition to a sphere
at the level of topology). 
Our approach  is  more rigorous than previous considerations ;
it seems that it is also simpler and more transparent. In particular, we 
obtain the ADHM equations in a very simple way. 
\end  {abstract}      											 
Gauge theories on noncommutative spaces introduced by A.Connes [1] 
play now an important role in string/M-theory. (Their appearance was 
understood first from the viewpoint of Matrix theory in [2]; see [3], [4]  
for review of results obtained in this framework  and [5] for the analysis 
in the framework of string theory.)   

Instantons in noncommutative gauge theories were introduced in [6] for
the case of noncommutative ${\bf R}^{4}$ and in [7] for the case of
noncommutative $T^{4}$. Later they were studied and applied in numerous
papers (see [8], [9] for review). In present paper we suggest another
approach
to instantons on noncommutative ${\bf  R}^4$. It seems that it makes the
theory much more transparent.

 In  commutative case one can study instantons directly on ${\bf  R}^4$;
then we should restrict ourselves to the gauge fields with finite
euclidean action (or, equivalently to fields that are gauge equivalent to
the trivial field at infinity). However, technically it is more convenient
to work on four-dimensional sphere $S^4$. 

 We will show that the situation in noncommutative case is quite similar.
The standard ${\bf  R}^4$ is conformally equivalent to $S^4$; therefore
the possibility to replace ${\bf  R}^4$ with $S^4$ in the study of
instantons is obvious. In noncommutative case we cannot apply this logic.
However, we will see that it is useful to consider a unital algebra
$\tilde {{\bf  R}}^4_{\theta}$ obtained from the algebra ${\bf
R}^4_{\theta}$ by means of addition of unit element. Here ${\bf
R}^4_{\theta}$ stands for the algebra of smooth functions on
noncommutative ${\bf  R}^4$ that tend to zero at infinity  
(multiplication is defined as a star-product.) As usual in
noncommutative geometry non-unital algebras are related to non-compact
spaces; adjoining  unit element corresponds to one-point compactification.
This means that the transition from ${\bf  R}^4_{\theta}$ to $\tilde {{\bf
R}}^4_{\theta}$ is analogous to  transition from ${\bf  R}^4$ to $S^4$ (at
the level of topology). 

 Therefore one can think that it is easier to study instantons and
solitons 
using $\tilde {{\bf  R}}^4_{\theta}$; we'll see that this is true. 

Physicists usually don't care of  the precise definition of the space of
functions 
they are working with (and usually this policy is completely justified). 
However, in the consideration of  noncommutative ${\bf R}^d$ it is 
important to be careful: different definitions of the space 
of functions on noncommutative space lead to different results. In 
the standard definition the algebra of functions on 
noncommutative  ${\bf R}^d$ is described as an algebra generated by 
noncommutative coordinates $\hat {x}^k$ obeying $[\hat {x}^k, 
\hat {x}^l]=i\theta ^{kl}$. We base our exposition of the theory of 
instantons on algebras  ${\bf R}_{\theta}^d$ and 
 $\tilde {{\bf R}}_{\theta}^d$ that don't contain $\hat {x}^k$ at 
all. The absense of accurate definition of an instanton was not harmless, 
it led to a confusion in the question about existence of 
$U(1)$-instantons. The present paper is not written as a rigorous 
mathematical paper, but it is based on accurate definitions (and it 
seems that the exposition can be made rigorous without essential 
difficulties).  

The paper is organized as follows.

 In Sec. 1 we consider instantons on  noncommutative ${\bf R}^4$ 
using the algebra $\tilde {{\bf R}}_{\theta}^4$; we show that the 
 noncommutative analog of ADHM data  arises very naturally in this 
approach. The construction of  noncommutative solitons also 
becomes simpler.
 In Sec.2 we show how to relate the consideration 
based  on the algebra ${\bf R}_{\theta}^4$ to the formalism of 
Sec. 1. 
In Sec. 3 we analyze the 
noncommutative analog of ADHM construction.
In Sec. 4 we  discuss some definitions and results of the  preceding
sections  in more 
general setting and more accurately. We also compare definitions 
of instantons on  noncommutative torus and on
noncommutative Euclidean space. 
An appendix written by
A. Connes contains a proof of the fact that   modules  used
 in present paper exhaust projective modules over appropriate
algebras. ( I could not find  such a proof in the literature;
 I am indebted to A. Connes for giving a proof.) 

Our paper does not depend on previous papers on  noncommutative
instantons, but we did not try to reproduce all known results using 
our approach. The necessary  definitions of notions of noncommutative 
geometry [10] are given (but sometimes we omit some details).
We use freely results of noncommutative geometry  without explicit 
references; most of results we need are contained in [1], [11].

{\bf Preliminaries.}

Let $A$ be an associative algebra. A vector space $E$ is a right 
$A$-module if we can multiply $e\in E$ by $a\in A$ from the right and 
this multiplication is distributive and associative (in particular, 
$(ea)b=e\cdot(ab)$ for $e\in E,\ a,b\in A$). Introducing the notation 
$\hat {a}e=ea$ we can say that to specify a right module we should 
assign to every $a\in A$ an operator $\hat {a}: E\rightarrow E$ in such 
a way that $\hat {ab}=\hat {b}\hat {a}$. A linear map $\varphi 
:E\rightarrow E$ where $E$ is an $A$-module is called an endomorphism 
if it is $A$-linear (i.e. $\varphi(ea)=\varphi(e)a$). The set of all 
endomorphisms can be considered as an algebra; it is denoted by ${\rm 
End}_AE$. We say that $A$ is an involutive algebra if it is equipped 
with  antilinear involution $a\rightarrow a^+$ obeying $(ab)^+=b^+a^+$.
A module $E$ over involutive algebra $A$ is a (pre)Hilbert module if it is 
equipped with $A$-valued inner product $<e_1,e_2>$ obeying 
$$<e_1a_1,e_2a_2>=a_1^+<e_1,e_2>a_2.$$
for all $e_1,e_2\in E,\  a_1,a_2\in A$. If  $E_1,E_2$ are two Hilbert 
$A$-modules we denote by $B(E_1,E_2)$ the set of all $A$-linear 
maps $T: E_1\rightarrow E_2$ having adjoints. (One says that 
$T^{\star}: E_2\rightarrow E_1$ is a map adjoint to $T$ if 
$<Tx,y>=<x,T^{\star}y>$.) Notice that it follows from this 
definition that the spaces $B(E_1,E_2)$ and $B(E_2,E_1)$ are 
complex conjugate.

In  a Hilbert module $E$ we can construct 
endomorphisms by the following formula 
$$\alpha (x)=\sum b_i <a_i,x>$$
where $x,a_i,b_i\in E$. These endomorphisms are called 
endomorphisms of finite rank.

The set $A^n$ consisting of column vectors with entries from $A$ 
can be considered as a Hilbert $A$-module with respect to inner 
product $$<a,b>=\sum a_i^+b_i.$$
In more invariant way we can define $A^n$ as a tensor product of 
$n$-dimensional vector space $V$ and the algebra $A$.

 If $A$ is a unital algebra the $A$-module $A^n$ is called a 
free module with $n$ generators. A projective module $E$ 
over a unital algebra $A$ is by definition a direct summand 
in $A^n$ . (We consider only finitely generated projective 
modules.)   Projective modules are   Hilbert  modules 
with respect to an inner  product inherited from $A^n$. 
All endomorphisms of a projective module $E$ have adjoints: 
${\rm End}_AE=B(E,E)$.

We gave all definitions for right modules, one can give similar 
definitions for left modules.

\centerline {{\bf Section 1.}}

Let us consider the algebra ${\bf  R}^d_{\theta}$ and the algebra $\tilde
{{\bf  R}}^d_{\theta}$ obtained from ${\bf  R}^d_{\theta}$ by means of
addition of unit element. Recall, that we consider elements of ${\bf
R}^d_{\theta}$ as smooth functions on ${\bf  R}^d$  tending  to zero at
infinity. (More precisely, $f\in {\bf R}_{\theta} ^d$ if   derivatives 
of all orders exist and tend to zero at infinity). The product of
functions 
$f,g\in{\bf R}^d_{\theta}$  (star-product)can be defined by the formula 
$$(f\star g)(x)=\int\int f(x+\theta u)g(x+v)e^{iu v}du dv.$$
Operators $\alpha_v:f(x)\rightarrow f(x+v)$ where $v\in {\bf  R}^d$
specify an action of  Lie group ${\bf  R}^d$  on ${\bf
R}^d_{\theta}$. The derivatives $\partial _i=\partial /\partial x^i$ are
infinitesimal automorphisms (derivations) of  ${\bf  R}^d_{\theta}$. If
$E$ is a (right) $A$-module where $A={\bf  R}^d_{\theta}$ then a
connection (a gauge field) on $E$ is specified by means of ${\bf
C}$-linear operators $\nabla_1,...,\nabla_d$ obeying the Leibniz rule:
$$\nabla _i(ea)=\nabla _ie\cdot a+e\cdot \partial _ia$$
where $e\in E,\  a\in A$.

 The same definition can be used in the case $A=\tilde {{\bf
R}}^d_{\theta}$ and in more general case when we have an action of Lie
algebra ${\bf  R}^d$ on algebra $A$. If  $\nabla ^{\prime}, \nabla$ are
two connections then the difference $\nabla_i^{\prime}-\nabla_i$ commutes
with multiplication by elements $a\in A$; in other words
$\nabla_i^{\prime}-\nabla_i$ is an $A$-linear map or an endomorphism of
the module $E$. It is convenient to fix a connection  $\nabla ^{(0)}$  on
$E$; then  every other connection can be represented in the form 
$$ \nabla _i=\nabla_i^{(0)}+\alpha_i ,$$
where $\alpha _i\in {\rm End}_AE$ are elements of the algebra ${\rm
End}_AE$ of endomorphisms of the module $E$. 

 Every algebra $A$ can be considered as a module over itself. More
precisely the multiplication from the right  gives $A$ a structure of a
right $A$-module that will be denoted by $A^1$; to obtain a left module we
should consider the  multiplication from the left. It follows from
associativity that the  multiplication from the left commutes with
multiplication from the right: $(ae)b=a(eb)$. This means that the
operators of left multiplication $\varphi_a: e\rightarrow ae$ are
endomorphisms of $A^1$ (of $A$ considered as a right module). If $A$ is a
unital algebra every endomorphism $\varphi : A^1\rightarrow A^1$ has this
form: $\varphi (e)=\varphi(1\cdot e)=\varphi(1)\cdot e$. This statement
can be applied to unital algebra  $\tilde {{\bf  R}}^d_{\theta}$, but for
the algebra   ${\bf  R}^d_{\theta}$ we have also other endomorphisms  of
$( {\bf  R}^d_{\theta})^1$. Namely, endomorphisms of  $({\bf
R}^d_{\theta})^1$ correspond to smooth  bounded functions on  
${\bf  R}^d $: every function $\alpha (x)$ of
this kind determines an endomorphism transforming $e(x)\in  ({\bf
R}^d_{\theta})^1$ into star-product $\alpha (x)\star e(x)$. 
(Talking about smooth bounded functions 
we have in mind functions having bounded derivatives 
$\partial _{\alpha}f$  where  $\alpha =(\alpha _1,...,\alpha_d )$
is a multiindex .)
 
 Derivatives $\partial _1,..,\partial _d$ satisfy the Leibniz rule and
therefore specify a connection on $A^1$ for $A= {\bf  R}^d_{\theta}$ or
$A=\tilde {{\bf  R}}^d_{\theta}$. All other connections have the form 
$$\nabla_ie(x)=\partial _ie(x)+\alpha_i(x)\star e(x)$$ 
where $\alpha_1(x),...,\alpha_d(x)$ are smooth functions that  are 
bounded  in the case of  ${\bf  R}^d_{\theta}$  and  tend
to a constant at infinity in the case of $\tilde {{\bf  R}}^d_{\theta} $.  

 A little bit more complicated module $A^n$ can be obtained if we take a
direct sum of $n$ copies of the module $A^1$. Elements of $A^n$ can be
considered as column vectors with entries from $A$. Endomorphisms of
$A^n$ can be identified with $n\times n$ matrices having endomorphisms of
$A^1$ as their entries. Every connection on $A^n=({\bf  R}^d_{\theta})^n$
or  $A^n=(\tilde {{\bf  R}}^d_{\theta})^n$ can be represented in the form 
\begin {equation}
\nabla _i e_k(x)=\partial _ie_k(x)+(\alpha _i(x))_k^l \star  e_l(x)
\end {equation}
where the entries of $n\times n$ matrices $\alpha _1(x),...,\alpha _d(x)$
obey the same conditions as functions $\alpha _i(x)$ in the expression for
the connection in $A^1$. 
 Algebras  ${\bf R}^d_{\theta}$ and  $\tilde {{\bf R}}^d_{\theta}$ are
involutive algebras with respect to complex conjugation. This means that
we can consider Hilbert modules over these algebras. We'll identify gauge
fields with unitary connections, i.e. connections represented by 
operators $\nabla _i$ obeying the condition
$$<\nabla _ia,b>+<a,\nabla _ib>=\partial_i<a,b>.$$

 Field strength of a gauge field
$F_{ij}=[\nabla_i,\nabla_j]$ obeys similar condition. 

 Modules we considered were Hilbert modules. We can construct also other
Hilbert modules in the following way. 

 Let us consider Hermitian operators $\hat {x}^k$  satisfying  
commutation relations 
\begin {equation}
 [\hat {x}^k, \hat {x}^l]=i\theta ^{kl}\cdot 1
\end {equation}
We will assume, that  $d=2n$ and  $\theta $  is a nondegenerate 
$2n\times 2n$ matrix; then these relations
are equivalent to canonical  commutation relations $[\hat {x}^k, \hat
{x}^{k+n}]=i,\ \  [\hat {x}^k, \hat {x}^l]=0$ if both $k,l\leq n$ or
$k,l>n$. (More precisely, we can write (2) in the canonical form replacing
$\hat {x}^i$  by their linear combinations.) It is well known (Stone-von
Neumann theorem) that irreducible representation of canonical  commutation
relations is unique and that one can obtain all other representations
taking direct sums of several copies of   irreducible representations. By 
means of Hermitian operators $\hat {x}^k$ specifying an irreducible 
representation of  (2) we can 
assign to every function  $\varphi(x)=\int\varphi(k)e^{ikx}dk $ 
an operator $\hat {\varphi}=\int\varphi(k)e^{ik\hat{x}}dk$. 
Taking $\varphi \in {\bf R}_{\theta}^d$ we obtain an  
${\bf R}_{\theta }^d$-module ${\cal F}$. Notice that for 
$f\in  S({\bf R}^n), \ \  \varphi\in {\bf R}_{\theta}^d$ we have 
$\hat {\varphi}f\in S({\bf R}^n)$ (see Sec. 4). This means that 
$S({\bf R}^n)$  can be regarded as an irreducible 
${\bf R}_{\theta}^d$-module; we will denote it by the same 
letter ${\cal F}$.
Every ${\bf R}^d_{\theta}$-module 
can be considered as   $\tilde {{\bf R}}^d_{\theta}$-module 
(the unit element is represented by identity operator), therefore 
we can regard ${\cal F}$ as an $\tilde {{\bf R}}^d_{\theta}$-module. 

The module ${\cal F}$ is a  projective 
 $\tilde {{\bf R}}^d_{\theta}$-module. 
This is clear from the following construction of it  that is 
more convenient for us. Let us fix a real function 
$p\in {\bf R}_{\theta}^d$ obeying $p\star p=p$ and $\int pdx=1$. (In the 
correspondence between elements of $ {\bf R}_{\theta}^d$ and operators 
such a function corresponds to one-dimensional projection $\hat {p}$.)
Then 
${\cal F}$ can be defined as a submodule of $\tilde {{\bf R}}_{\theta}^d$ 
consisting of elements of the form $p\star r$ where $r\in  \tilde 
{{\bf R}}_{\theta}^d$. (If $\hat {p}$ is a projection on one-dimensional 
space spanned by a normalized vector $\alpha$ matrix elements of $\hat
{p}$ 
are given by the formula $<x\mid p\mid x^{\prime}>=\bar {\alpha}(x) 
\alpha (x^{\prime})$ and 
matrix elements of the operator corresponding to $p\star r$ have the form 
$\bar {\alpha}(x)\rho (x^{\prime})$ for some function $\rho\in S({\bf
R}^n)$.) The free 
module $ \tilde {{\bf R}}_{\theta}^d$ is a direct sum of a module ${\cal
F}$ and 
a module consisting of elements of the form $(1-p)\star r$ therefore 
${\cal F}$ is projective. It follows from this remark that the module
${\cal F}$ 
can be characterized also as a submodule consisting of elements $v$
obeying 
$(1-p)\star v=0$.  Replacing the condition $\int pdx=1$ by the 
condition $\int pdx=k\in {\bf Z}$ we obtain by means of the same 
construction a module isomorphic to ${\cal F}^k$. (The  corresponding 
operator $\hat {p}$ is a projection on a $k$-dimensional space $E_k$
and has matrix elements of the form $<x\mid \hat {p}\mid x^{\prime}>= 
\bar {\alpha}_1 (x)\alpha_1 (x^{\prime})+...+\bar {\alpha} _k(x) 
\alpha _k(x^{\prime})$ where $\alpha _1,...,\alpha_k$ stands for 
orthonormal basis of $E_k$.
Matrix elements of the operator corresponding to $p\star r$ have the form 
$\bar {\alpha}_1 (x)\rho_1 (x^{\prime})+...+\bar {\alpha} _k(x)\rho
_k(x^{\prime})$ for 
some functions  $\rho_1,...,\rho_k$.)

Similar statement is true for projections on $ ({\bf R}_{\theta}^d)^s$. 
Considering $\int pdx$ as a trace on $ {\bf R}_{\theta}^d$ we define a 
trace on matrices with entries from $ {\bf R}_{\theta}^d$. If $\Pi $ is a 
self-adjoint $s\times s$ matrix of this kind, $\Pi ^2 =\Pi $ and  ${\rm
Tr} 
\Pi =k$ then the matrices of the form $\Pi R$, where $R$ is an $s\times s$ 
matrix with entries from $\tilde {{\bf R}}_{\theta}^d$ constitute an 
$\tilde {{\bf R}}_{\theta}^d$-module isomorphic to ${\cal F}^k$.
Here multiplication of matrices is understood as a combination of the 
usual matrix product and the star-product. A matrix $V=\Pi R$ obeys 
$(1-\Pi )V=0$. This remark leads to an alternative definition of the
module 
at hand: considering $1-\Pi$ as an operator acting on 
$({\bf R}_{\theta}^d)^s$ we can say that ${\rm  Ker}(1-\Pi)$ is 
isomorphic to ${\cal F}^k$.

Operators $\nabla_k^{(0)}$ related with $\hat {x}^l$ by the formula $\hat
{x}^k=i\theta^{kl}\nabla_l^{(0)}$  specify a connection on the module
${\cal F}$. It follows from Schur's lemma that all  endomorphisms of
${\cal F}$ have the form ${\rm const}\cdot 1$; therefore all other
connections can be represented as $\nabla_k=\nabla_k^{(0)}+c_k\cdot 1$.
We'll  study  $\tilde {{\bf R}}^d_{\theta}$-modules ${\cal F}_{rs}$ 
that are represented as direct 
sums of $r$ copies of the module ${\cal F}$ and $s$ copies of the module
$(\tilde {{\bf R}}^d_{\theta})^1$. All these modules are projective. One 
can prove that every projective $\tilde {{\bf R}}^d_{\theta}$-module is 
isomorphic to one of modules  ${\cal F}_{rs}$  ( see Sec.4 and Appendix). 

 Every connection on ${\cal F}_{rs}$ can be written in the form 
\begin {equation}
\nabla _k=\nabla _k^{(0)}+\Phi ,\ \  \Phi = \left ( \begin {array} {cc}
                                                            M_k & N_k \\
                                                            S_k & T_k
 \end {array}        \right )
\end {equation}
where  $\nabla_k^{(0)}$ is
 the standard  connection acting as $i(\theta
^{-1})_{kl}\hat {x}^l$ on ${\cal F}$ and as $\partial _k$ on   $({\bf
R}^d_{\theta})^1$,   $M_k$ is an $r\times r$ matrix with entries from
${\bf C}$,   $N_k$ is an $r\times s$ matrix with entries from ${\cal F},
S_k$ is an $s\times r$ matrix with entries from $\bar {{\cal F}}$ 
and $T_k$ is an $s\times s$ matrix with entries from 
$\tilde {{\bf R}}^d_{\theta}$. (Here $\bar {{\cal F}}$ is obtained 
from ${\cal F}$ by means of complex conjugation.)  This 
follows from the remark that an endomorphism of direct sum 
$E_1+...+E_r$ where $E_i$ are $A$-modules can be 
described as a matrix having an $A$-linear map from $E_k$ in 
$E_l$ as an element in
$k$-th row and $l$-th column. (For any  unital algebra $A$ an 
$A$-linear map from  $A^1$ into any $A$-module $E$ is 
characterized by the image of
unit element; this means that the space of $A$-linear maps $A^1\rightarrow
E$ can be identified with $E$. The space of $\tilde {{\bf
R}}^d_{\theta}$-linear 
maps from ${\cal F}$ into $\tilde {{\bf R}}^d_{\theta}$  can be identified
with 
$B({\cal F},\tilde {{\bf R}}^d_{\theta})$; it is complex conjugate to 
$B(\tilde {{\bf R}}^d_{\theta},{\cal F})={\cal F}$.

It is easy to see that the connection $\nabla_k^{(0)}$ satisfies the
Yang-Mills 
equations of motion. Let us consider Yang-Mills field interacting with
scalar 
field $\varphi$ in adjoint representation; if Yang-Mills field is regarded
as a 
connection on a module ${\cal F}_{rs}$  then $\varphi$ should be regarded 
as an endomorphism of this module. We denote by $\varphi _0$   an 
endomorphism of  ${\cal F}_{rs}$ acting as an identity map on ${\cal F}^r$ 
and as a zero map on  $\tilde {{\bf R}}^s_{\theta}$. Taking the gauge
field as 
$\nabla _k^{(0)}$ and the scalar field as $\alpha \varphi _0$  we obtain a 
solution to the equations of motion   if $\alpha$ and zero are  stationary 
point of the potential of the scalar field . It is easy to check that this
simple 
solution corresponds to the solution of [13]; see Sec 2.  This is an
additional 
confirmation of the idea that it is necessary to use not only free
modules, 
but also other modules;  a consideration of classical solutions to 
equations of motion based on this idea will be given in [14].

 Now we can turn to the study of instantons on    $\tilde {{\bf
R}}^4_{\theta}$-modules ${\cal F}_{rs}$. By definition, an instanton is a
gauge field (unitary connection) obeying $F^+=0$ where $F^+$
is the self-dual part of the curvature $F_{kl}=[\nabla_k,\nabla_l]$. It is
convenient to introduce operators 
\begin {equation}
\begin {array} {cc}
D_1=\nabla_1+i\nabla _2, & D_2=\nabla_3+i\nabla _4,  \\
D_1^+=-\nabla_1+i\nabla _2, & D_2^+=\nabla_3+i\nabla _4
\end {array}
\end {equation}
Then an instanton satisfies 
\begin {equation}
[D_1,D_2]  = 0,  
\end {equation}
\begin {equation}
[D_1,D_1^+]  +  [D_2,D_2^+]=0.
\end {equation}
Let us represent $D_1$ and $D_2$ in the form 
\begin {equation}
D_1=D_1^{(0)}+\left ( \begin {array}{cc}
B_1 & I \\
K  & R_1
\end {array}  \right ) ,\ \ D_2=D_2^{(0)}+\left ( \begin {array} {cc}
B_2 & L \\
J & R_2
\end {array} \right ).
\end {equation}
We'll assume that $K=0$ and $L=0$ (these conditions can be considered as
gauge conditions).  Then the equation (5) can be rewritten in the
following
way 
\begin {equation}
\alpha \cdot 1 +[B_1, B_2]+IJ=0,
\end {equation}
\begin {equation}
IR_2-B_2I-D_2^{(0)}I=0,
\end {equation}
\begin {equation}
R_1J-JB_1+D_1^{(0)}J=0,
\end {equation}
\begin {equation}
-JI+[R_1,R_2]+\partial _1^{(0)}R_2-\partial _2^{(0)}R_1=0.
\end {equation}
Multiplication  in these formulas is considered as the usual matrix
multiplication combined with natural bilinear maps ${\cal F}\otimes \bar
{{\cal F}}\rightarrow {\bf C}, \ \  {\cal F}\otimes  \tilde {{\bf
R}}^d_{\theta}\rightarrow {\cal F},\ \   \tilde {{\bf
R}}^d_{\theta}\otimes \bar {{\cal F}}\rightarrow \bar {{\cal F}}, \ \
\tilde {{\bf R}}^d_{\theta} \otimes  \tilde {{\bf R}}^d_{\theta}
\rightarrow  \tilde {{\bf R}}^d_{\theta} , \ \  \bar {{\cal F}}\otimes
{\cal F}\rightarrow  \tilde {{\bf R}}^d_{\theta}$. We use these maps for
$d=4$, but they are defined for every even $d$ and nondegenerate $\theta$.
The first of these maps stems from hermitian inner product, the second
from action of $ \tilde {{\bf R}}^d_{\theta}$ on ${\cal F}$. The last map
is an  $\tilde {{\bf R}}^d_{\theta}$-valued inner product on ${\cal F}$ 
(the module ${\cal F}$ is projective and therefore can be considered as 
a Hilbert module).
The operators $D_i^{(0)}, \partial _i^{(0)},\ \  i=1,2$ are related to
$\nabla
_i^{(0)},\partial _i,\ \  i=1,2,3,4$ in the same way as $D_1 ,D_2$ are
related 
to $\nabla_i$. One can express $\alpha \cdot 1=[D_1^{(0)},D_2^{(0)}]$ in 
terms  of $\theta ^{jk}$ (namely, $\alpha=\theta ^{13}-\theta
^{24}+i\theta^{23}+i\theta^{14}$).  The equation (6) also can be
represented by equations for four blocks; we write down only the first
one:  
\begin {equation}
\beta \cdot 1+[B_1,B_1^+]+[B_2,B_2^+]+II^+-J^+J=0
\end {equation}
where $\beta \cdot 1=[D_1^{(0)},D_1^{(0)+}]+[D_2^{(0)},D_2^{(0)}]^+$.  It
is easy to see that equations (8) and (12) are closely related to ADHM
construction; moreover, they coincide with noncommutative counterpart of
ADHM equations if we make an ansatz $I=I^{(0)}\cdot \Phi, \ \
I=J^{(0)}\cdot  \Psi$ where $I^{(0)}$ and $J^{(0)}$ are matrices with
complex entries, $\Phi \in {\cal F},\ \  \Psi \in \bar {{\cal F}}, \ \
\Phi \Psi =1$. It is natural  to conjecture that having a solution of
noncommutative ADHM equations (8), (12)  we can find $\Phi ,\Psi ,R_1,
R_2$ 
in such a way that (5), (6) are satisfied.

The above consideration can be applied with minor modifications 
to instantons on noncommutative orbifold ${\bf R}_{\theta}^4 /\Gamma$; 
it leads to noncommutative analog of equivariant ADHM equations found in
[12].

\centerline {{\bf Section 2.}}

In this section we'll 
consider instantons on noncommutative ${\bf R}^4$ 
in terms of the algebra ${\bf R}_{\theta}^4$.  We'll relate 
this consideration to the analysis in terms of 
$\tilde {{\bf R}}_{\theta}^4$ given in Sec. 1.

Let us recall the standard definition of instanton on commutative  
${\bf R}^4$. In this definition we consider solutions of 
(anti)selfduality equation, that are gauge trivial at infinity. 
More precisely we restrict ourselves to gauge fields that 
can be represented at infinity in the form 
\begin {equation}
A_{\mu}\approx g ^{-1} (x) \partial _{\mu}g(x)
\end {equation}
where $g(x)$ is a function taking values in the gauge group $G$. 
The function $g(x)$ is defined outside some ball $D$ and cannot 
be extended to the whole space ${\bf R}^4$ (the obstruction to this 
extension can be identified with topological number of the gauge 
field). To find an appropriate condition of "triviality at infinity" 
on noncommutative ${\bf R}^4$ we start with reformulation of this 
condition in the commutative case. We assume that $G=U(n)$ and 
consider matrix valued functions $g(x),\  h(x)$ satisfying the 
condition $$h(x)g(x)=1-p(x)$$ where $p(x)$ rapidly tends to 
zero at infinity. Then a gauge field $A_{\mu}$ is trivial at 
infinity if it can be represented in the form
\begin {equation}
A_{\mu}(x)=h(x)\partial _{\mu}g(x)+\alpha_{\mu}(x)
\end {equation}  
where $\alpha _{\mu}$  tends to zero at infinity faster  than 
$\parallel x\parallel ^{-1}$.  (If  $A_{\mu}$ is represented in the 
form (13) we can obtain a   representation in the form (14) taking 
as $g(x),\  h(x)$ 
extensions of $g(x),\  g^{-1}(x)$  to the whole space ${\bf R}^4$.  
Such extensions don't exist if we consider $g(x),\  g^{-1}(x)$ 
as $U(n)$-valued functions, but they do exist if we consider them 
as matrix valued functions.)

 It is convenient to represent (14) in the form 
\begin {equation}
\nabla _{\mu}=\hat {h}\circ \partial _{\mu}\circ \hat {g} 
+\hat {\beta}_{\mu}
\end {equation}
where $\hat {h}, \hat {g}, \hat {\beta}_{\mu}$ stand for 
operators of multiplication by $h,g,\beta _{\mu}$. We'll 
use this representation to generalize the above considerations 
for the gauge fields on noncommutative ${\bf R}^4$. 
We consider these gauge fields as connections on the module 
$({\bf R}_{\theta}^4)^n$. We have seen that such connections 
can be represented in the form (1), where $\alpha _i$ belong 
to the algebra 
$B={\rm End}_{{\bf R}_{\theta}^4}({\bf R}_{\theta}^4)^n$ of 
endomorphisms of $({\bf R}_{\theta}^4)^n$ (they can be 
considered as $n\times n$ matrices having bounded smooth functions 
 as their entries). 

Let us consider the subalgebra $B_0\subset B$ consisting 
of all endomorphisms tending to zero at infinity faster 
than $\parallel x\parallel ^{-1}$. More precisely, 
elements of $B_0$ should be represented 
by matrices with entries belonging to the algebra 
 $\Gamma _{\rho}^m$ with $m<-1$. (See Sec.4 for the 
definition of $\Gamma _{\rho}^m$.) 
 This means that $B_0$ can be 
regarded as a subalgebra of the algebra of endomorphisms 
of the $\tilde {{\cal A}}$-module $(\tilde {{\cal A}})^n$ 
 where we can take ${\cal A}={\bf R}_{\theta}^4$  or  
${\cal A}=\Gamma _{\rho}^m$, $m<-1$.

Let us say that a gauge field $\nabla _{\mu}$ on 
${\bf R}_{\theta}^4$ is gauge trivial at infinity if 
\begin {equation}
\nabla _{\mu}=T\circ \partial _{\mu}\circ T^++ (1-T T^+)\circ 
\partial _{\mu}+\sigma _{\mu}
\end {equation}
where $T^+T=1,\  \sigma_{\mu}\in B_0$ and $T^+$ is a parametrix 
of $T$, i.e. $1-TT^+=\Pi$ is a matrix with entries from the 
Schwartz space $S({\bf R}^4)$. Here $T\in B 
={\rm End} _{{\bf R}_{\theta}^4}({\bf R}_{\theta}^4)^n$ is an 
$n\times n$ matrix considered as an endomorphism of 
$({\bf R}_{\theta}^4)^n$. (See Sec. 4 for the conditions of 
existence of parametrix.)

Having an endomorphism $T$ obeying the above conditions 
we can show that ${\rm Ker} T^+\oplus  
({\bf R}_{\theta}^4)^n$ is 
isomorphic to $({\bf R}_{\theta}^4)^n$ as a Hilbert 
${\bf R}_{\theta}^4$-module.
The construction of the isomorphism goes as follows. We map 
$(\xi ,x)\in {\rm Ker}T^+\oplus ({\bf R}_{\theta}^4)^n$ into 
$\xi +Tx\in ({\bf R}_{\theta}^4)^n$. The inverse map transforms 
$y\in ({\bf R}_{\theta}^4)^n$ into $(\Pi y,T^+y)$. (The map 
$\Pi=1-TT^+$ is a projector of $({\bf R}_{\theta}^4)^n$ onto 
${\rm Ker}T^+$). These maps preserve ${\bf R}_{\theta}^4$-valued 
inner products: $<\xi +Tx, \xi^{\prime}+Tx^{\prime}>= 
<\xi ,\xi^{\prime}>+<Tx, \xi^{\prime}>+<\xi  , Tx^{\prime}> 
<Tx,Tx^{\prime}>=<\xi ,\xi^{\prime}>+<x, T \xi ^{\prime}>+ 
<T^+\xi ,x^{\prime}>+<T^+Tx, x^{\prime}>=<\xi,\xi^{\prime}> 
+<x,x^{\prime}>$. It follows from $\Pi\in S({\bf R}^4)$ that the  
${\bf R}_{\theta}^4$-module ${\rm Ker}T^+$  is isomorphic 
to ${\cal F}^k$ for  same $k$. (We are using the fact that 
${\rm Ker}TT^+={\rm Ker} T^+$ and that 
${\rm Ker}(1-\Pi)={\rm Ker}TT^+$ is isomorphic to ${\cal F}^k$.)                        
We can conclude therefore that ${\cal F}^k\otimes ({\bf R}_{\theta}^4)^n$ 
is isomorphic to $({\bf R}_{\theta}^4)^n$ as Hilbert ${\bf
R}_{\theta}^4$-module.

Let us analyze the relation between connections on modules 
$({\bf R}_{\theta}^4)^n$  and ${\rm Ker} T^+\oplus ({\bf
R}_{\theta}^4)^n$.

The trivial connection $\partial _{\alpha}$ on $({\bf R}_{\theta}^4)^n$  
induces a connection  $\Pi\partial _{\alpha} \Pi$ on ${\rm Ker}T^+$. Let
us 
calculate the connection $D_{\alpha}$ on  $({\bf R}_{\theta}^4)^n$ that    
corresponds to the connection $(\Pi\partial _{\alpha}\Pi, \partial
_{\alpha})$ on 
${\rm Ker} T^+\oplus ({\bf R}_{\theta}^4)^n$. We obtain $y\rightarrow 
(\Pi y, T^+y)\rightarrow (\Pi\partial _{\alpha}\Pi y, \partial
_{\alpha}T^+ 
y)\rightarrow  \Pi\partial _{\alpha}\Pi y+T\partial _{\alpha}T^+$ hence
\begin {equation} 
D_{\alpha}=T\partial _{\alpha} T^++\Pi\partial _{\alpha}\Pi=T_{\alpha} 
\partial _{\alpha}T^++\Pi\partial _{\alpha}+\rho _{\alpha} 
\end {equation}
where $\rho _{\alpha}=\Pi[\partial _{\alpha},\Pi]$. We see that
$D_{\alpha}$
is gauge trivial at infinity. Gauge 
fields on  $({\bf R}_{\theta}^4)^n$  that are gauge trivial at infinity
(i.e. have 
the form (16)) correspond on  ${\rm Ker }T^+\oplus ({\bf R}_{\theta}^4)^n$ 
to gauge fields of the form 
\begin {equation}  
(\Pi\partial _{\alpha}\Pi ,\partial _{\alpha})+\nu _{\alpha}
\end {equation}
where $\nu  _{\alpha}$ have the same behavior at infinity as 
$\sigma _{\alpha}$ in (16). More precisely, if $\sigma _{\alpha} 
\in \Gamma _{\rho}^m$ then the field (18) can be extended  to 
$\tilde {\Gamma}_{\rho}^m$-module ${\cal F}_{kn}$ (and, because 
$\Gamma _{\rho}^m\subset {\bf R}_{\theta}^4$, also to ${\cal F}_{kn}$ 
considered  as $\tilde {{\bf R}}_{\theta}^4$-module).
 
It is clear that in the  above consideration we can replace 
 ${\bf R}_{\theta}^4$ with    ${\bf R}_{\theta}^d$. (The matrix $\theta $ 
should be nondegenerate, hence $d$ is necessarily even.)  Endomorphisms  
obeying $T^+T=1$ (isometries) were used in several papers to construct  
noncommutative  solitons (see, for example, [8], [9], [13], [16], [17]).
The approach 
of these papers is closely related to our treatment. In particular, the 
simple solution to the equation of motion for Yang-Mills field 
interacting with scalar field described in Sec. 1 can be transformed into 
noncommutative soliton of [13] by means of the isomorphism between 
${\rm Ker}T^+\otimes ({\bf R}_{\theta}^{2k})^n$ 
and  $({\bf R}_{\theta}^{2k})^n $.  The standard  connection 
$\nabla _{\alpha}^{(0)}$ of Sec.1 is an instanton if the matrix 
$\theta _{\alpha \beta}$ is antiselfdual; it corresponds to the 
instanton found in [17].
Now we are able to compare different definitions of instantons. One can 
consider instantons as (anti)selfdual  gauge fields  on  
$({\bf R}_{\theta}^4)^n$  that are  gauge trivial  at infinity. It follows
from 
the above consideration that equivalently we can consider instantons as
gauge 
fields (18) on ${\cal F}^k\oplus ({\bf R}_{\theta}^4)^n$  obeying 
(anti)selfduality equation. Using the fact that fields (18) can be
extended to  
$\tilde {{\bf R}}_{\theta}^4$-module ${\cal F}_{kn}$ we obtain  the
relation 
between instantons on $({\bf R}_{\theta}^4)^n$ and instantons studied  
in Sec. 1. Namely, we see that instantons on ${\cal F}_{rs}$ (the 
fields (7) obeying equations (5),(6)) correspond to instantons on 
$({\bf R}_{\theta}^4)^s$ if $R_1, R_2$ are matrices with entries 
from $\Gamma_{\rho}^m$, $m<-1$.

\centerline {{\bf Section 3.}}
 
Let us
analyze the noncommutative analog of ADHM construction [1], [3]  
in our approach. We consider an ${\bf R}_{\theta}^4$-linear operator 
$$D^+:(V\oplus V\oplus W)\otimes {\bf R}_{\theta}^4\rightarrow 
(V\oplus V)\otimes {\bf R}_{\theta}^4 $$
defined by the formula 
\begin {equation}
D^+=\left ( \begin {array}{ccc}
-B_2+z_2 & B_1-z_1 & I \\
B_1^+-\bar {z}_1 & B_2^+-\bar {z}_2 & J^+
\end {array} \right ).
\end {equation}
Here $B_1, B_2\in {\rm Hom}(V,V),\  I\in {\rm Hom} (W,V),\  J\in 
{\rm Hom} (V,W),\  z_1=\hat {x}^1+i\hat {x}^2,\  z_2=\hat {x}^3+ 
i\hat {x}^4,\  [x^r,x^s]=i\theta ^{rs}$.
Commutation relations between $z_i, \bar{z}_i$ can be written in 
the form 
$$[z_1,z_2]=-\zeta_c,\  [z_1,\bar {z}_1]=-\zeta_1,\  
[z_2,\bar{z}_2]=-\zeta_2 $$
(We use the notations of [3].)
We impose the condition that the operator 
$$D^+D:(V\oplus V)\otimes {\bf R}_{\theta}^4 \rightarrow (V\oplus V) 
\otimes  {\bf R}_{\theta}^4$$
has the form 
\begin {equation}
\left ( \begin {array}{cc}
\Delta & 0 \\
0 & \Delta
\end {array} \right )
\end {equation} 
where $\Delta :V\otimes {\bf R}_{\theta}^4 \rightarrow V\otimes 
{\bf R}_{\theta}^4$  can be written as 
$$\Delta =(B_1-z_1)(B_2^+-\bar {z}_1)+(B_2-z_2)(B_2^+-\bar {z}_2) 
+II^+$$
$$=(B_1^+-\bar {z}_1)(B_1-z_1)+(B_2^+-\bar {z}_2)(B_2-z_2)+J^+J.$$

It is easy to check that this condition is equivalent to the
noncommutative 
ADHM equations (8), (12). One can check that the operator $\Delta $ 
has no zero modes (see [3]). 

Let us analyze the space ${\cal E}\subset 
(V\oplus V\oplus W)\otimes {\bf R}_{\theta}^4$ consisting of solutions 
to the equation $D^+\psi =0$. From ${\bf R}_{\theta}^4$-linearity of 
$D^+$ it follows that ${\cal E}$ can be considered as ${{\bf R}} 
_{\theta }^4$-module.

One can consider an orthogonal projection $P$ of the module 
$(V\oplus V\oplus W)\otimes {\bf R}_{\theta}^4$  onto ${\cal E}$;
the standard connection $\partial _{\alpha}$ on 
$(V\oplus V\oplus W)\otimes {\bf R}_{\theta}^4$ generates an instanton 
$P\circ \partial _{\alpha}\circ P$ on ${\cal E}$ (see[1], [3]).  An 
equivalent way to 
construct an instanton is to consider an isometric ${\bf R}_{\theta}^d$-
linear map  $\omega :{\cal E}^{\prime}\rightarrow (V\oplus V\oplus 
W)\otimes {\bf R}_{\theta}^4$ having ${\cal E}$ as its image and to 
define connection on ${\cal E}^{\prime}$ by the formula 
$\omega^+\circ \partial_{\alpha}\circ \omega$. (Here 
${\cal E}^{\prime}$ is a Hilbert module, a map is isometric if 
$\omega^ +\omega=1$.)

We will relate this construction to the above consideration finding 
explicitly the module ${\cal E}^{\prime}$ and the map $\omega$. 
This will allow us to say that gauge fields obtained by means of 
noncommutative ADHM construction are instantons in the sense of present
paper. 

Representing $\psi$ as a column vector 
\begin {equation}
\left( \begin {array} {c}
\varphi \\
\zeta
\end {array}  \right)
=
\left( \begin {array} {c}
\varphi_1 \\
\varphi_2\\
\zeta
\end {array}  \right)
\end {equation}
 where $\varphi_i\in V\otimes  {\bf R}_{\theta}^4, \  
\xi \in W\otimes {\bf R}_{\theta}^4$ 
we can rewrite the equation $D_+\psi=0$ in the form 
\begin {equation}
A\varphi +C\xi=0 
\end {equation}
where 
\begin {equation}
A\varphi=\left( \begin {array} {cc}
-B_2+z_2 & B_1-z_1\\
B_1^+-\bar{z}_1 & B_2^+-\bar {z}_2
\end {array}     \right) 
\left( \begin {array} {c}
\varphi _1\\
\varphi _2
\end {array}  \right)  ,\ \ \  C\zeta =
\left( \begin {array}{c}
I\xi\\
J^+\xi
\end {array}  \right)
\end {equation}
We will assume that $\theta$ is a nondegenerate matrix; 
then without loss of generality we can assume that 
$\zeta _1>0, \  \zeta_2>0$. Our assumption allows us to consider 
elements of ${\bf R}_{\theta}^4$ as pseudodifferential 
operators acting on functions defined on ${\bf R}^2$. Then 
all above equations can be regarded as operator equations. In 
particular, $A$ is a first order differential elliptic operator of index 
$r={\rm dim} V$ and has a parametrix $Q$, a pseudodifferential 
operator obeying $QA=1-\Pi ,\  AQ=1-\Pi^{\prime}$ where 
$\Pi,\Pi^{\prime}$ are operators of finite rank. 

The operator equation (22) can be reduced to the equation 
\begin {equation}
Af+Cg=0
\end {equation}
where $f$ and $g$ are functions defined on ${\bf R}^2$. If 
${\cal N}$ is the space of solutions to (24) then solutions to (22) 
can be characterised as pseudodifferential operators taking values 
in ${\cal N}$. Using the parametrix we can reduce the study  of
Eqn (24) to the analysis of equation $(1-\Pi )f+QCg=0$; this is 
essentially a finite-dimensional problem. We will assume that 
$A\cdot Q=1$ (i. e. $\Pi^{\prime}=0$); this technical 
assumption leads to some simplication of calculations. In our 
conditions the operator $A$ has a kernel ${\rm Ker }A$ of 
dimension $r={\rm dim}V$. Every solution of (24) can be written as 
$F-QCg$ where $F\in {\rm Ker}A$. This permits us to say that the 
 ${\bf R}_{\theta}^4$-module ${\cal E}$ is isomorphic to the 
direct sum  
$${\cal E}^{\prime}={\cal F}^r\otimes (W\otimes {\bf R}_{\theta}^4).$$
An ${\bf R}_{\theta}^4$-linear map $v:{\cal E}^{\prime}\rightarrow 
(V\oplus V\oplus W)\otimes {\bf R}_{\theta}^4 $ defined by the 
formula
\begin {equation}
v
\left( \begin {array} {c}
\rho \\
\xi
\end {array}  \right)   =
\left( \begin {array}{c}
\rho-QC\xi \\
\xi
\end {array}  \right)
\end {equation}
is an isomorphism between ${\cal E}^{\prime}$ and ${\cal E}$. (We 
consider ${\cal F}^r$ as a space consisting of  pseudodifferential 
operators $\rho \in (V\oplus V)\otimes {\bf R}_{\theta}^4$ taking values 
in ${\rm Ker}A$; if $f_1,...,f_r$ is a basis of ${\rm Ker } A$ then 
$<x\mid \rho \mid x^{\prime}>=\sum_{1\leq k \leq r} 
f_k(x)\rho_k(x^{\prime})$.) An isometric map $w: {\cal E}^{\prime} 
\rightarrow {\cal E}$ can be constructed as $w=v(v^+v)^{-1/2}$.

   To prove that the map $v$ is an isomorphism and to check that the 
gauge field $\omega ^+\circ\partial _j\circ \omega$ is an instanton 
we need some information that can be obtained easily from well known 
results about  pseudodifferential operators [15]; necessary results 
are formulated in Sec.4. Notice that at infinity $\omega$ has the same
behavior as $v$. This follows from the remark that $v^+v-1$ tends to zero
at infinity as 
$\parallel z\parallel ^{-1}$. (Using the notations of Sec.4 
we have $A\in H\Gamma_1^{1,1}$, $C\in \Gamma_1^0$, $Q\in
H\Gamma_1^{-1,-1}$, hence $v^+v-1\in \Gamma_1^{-1}$.)This 
remark permits us to say that the asymptotic behavior of 
$\omega ^+\circ \partial \circ \omega $ is the same as 
the asymptotic behavior of $v ^+\circ \partial \circ v$. We 
obtain that the difference between  
$\omega ^+\circ \partial \circ \omega $ and the standard 
connection on ${\cal E}^{\prime}$ tends to zero as 
$\parallel z\parallel ^{-2}$. (More precisely the entries 
of corresponding matrices belong to $\Gamma_1^{-2}$.)

 \centerline {{\bf Section 4.}}

In this section we'll describe some general properties of instantons on 
noncommutative spaces. We'll use in our consideration a general 
construction of deformations of associative algebra equipped with an 
action of commutative Lie group; this construction was studied in [11].
We will list some properties of the star-product that can be obtained 
by the methods of the theory of  pseudodifferential operators. These 
mathematical results permit us to justify some statement of 
preceding sections. 

 Let us consider an associative algebra $A$ and a $d$-dimensional abelian
group $L={\bf R}^d$ acting on $A$ by means of automorphisms. This means
that for every $v\in {\bf R}^d$ we have an   automorphism  $\alpha
_v:A\rightarrow A$ and $\alpha_{v_1+v_2}=\alpha_{v_1} \cdot \alpha_{v_2}$.
We assume that $A$ has a norm $\parallel \ \  \parallel$ and an involution
$^\ast $ that are invariant with respect to  automorphisms  $\alpha _v$. 
It is assumed also that the automorphisms $\alpha _v$ are strongly
continuous 
with respect to $v$.

One says that an element $a\in A$ is smooth if the function $v\rightarrow
\alpha _v(a)$ is infinitely differentiable $A$-valued function on ${\bf
R}^d$ (differentiation is defined with respect to the norm   $\parallel \
\  \parallel$ on $A$). The set $A^{\infty}$ of all smooth elements
constitutes a subalgebra of $A$; the Lie algebra ${\cal L}$ of the group
$L$ 
acts on $A^{\infty}$ by means of infinitesimal  automorphisms
(derivations). 

 If $\theta$ is a $d\times d$ antisymmetric matrix we can introduce a new  
product in $A^{\infty}$ (star-product) by means of the following formula:
\begin {equation}
f\star g=\int \alpha_{\theta  u}f\cdot \alpha_vg\cdot e^{iuv}dudv.
\end {equation}
This new operation is also an associative product and the action of the
group $L={\bf R}^d$ preserves it [11]. It depends continuosly on $\theta$;
therefore the new algebra $A_{\theta }^{\infty}$ (the algebra of smooth
elements with respect to the product (26)) can be regarded as a
deformation
of the algebra $A^{\infty}$.

  Let us consider some examples. One can take as $A$ the algebra $C(T^d)$ 
of continuous functions on a torus $T^d={\bf R}^d/{\bf Z}^d$ equipped with 
the supremum norm. The Lie group $L={\bf R}^d$ acts on $T^d$ in natural 
way  (by means of translations) and this action generates  an action 
of $L={\bf R}^d$ on $A=C(T^d)$. The algebra $A^{\infty}$ consists of 
smooth functions on $T^d$. The star-product coincides with Moyal 
product as in all of our examples and $A_{\theta}^{\infty}$ is 
interpreted as an algebra of smooth functions on noncommutative 
torus; we will denote it by $T_{\theta}$. 

Let us take as $A$ the algebra $C_0({\bf R}^d)$ of   continuous functions 
on ${\bf R}^d$ that tend to zero at infinity (the norm is again the 
supremum norm). Using the natural action of $L={\bf R}^d$ on $A$ we obtain 
an algebra ${\bf R}_{\theta}^d$ that can be  interpreted as an algebra of 
smooth functions on noncommutative Euclidean space tending  to zero at 
infinity. We can consider instead of $C_0({\bf R}^d)$ the algebra 
$C({\bf R}^d \cup \infty)$ of continuous functions on ${\bf R}^d$, that  
tend to a constant  at infinity. Every element of this algebra can be 
represented in the form $f+\alpha \cdot 1$ where $f\in C_0({\bf R}^d)$. 
For any algebra $A$ we denote by $\tilde {A}$ the algebra consisting 
of elements of the form $a+\alpha \cdot 1$ where $a\in A$, $\alpha \in 
{\bf C}$ and $1$ is a unit element: $a\cdot 1=1\cdot a=a$. (One says that 
$\tilde {A}$ is a unitized algebra $A$.) We see that $C({\bf R}^d\cup 
\infty)=\widetilde {C({\bf R}^d)}$. The Lie group $L={\bf R}^d$ acts on 
$\widetilde {C({\bf R}^d)}$. Applying the above construction to this
action 
we obtain a unital algebra $\tilde {{\bf R}}_{\theta }^d$; one can get it 
adding unit element to the algebra 
${\bf R}_{\theta}^d$ of the preceding example.
Starting with the algebra of all bounded uniformly continuous functions on 
${\bf R}^d$  with action of  translation group  $L={\bf R}^d$  we obtain
an algebra of  bounded functions that  have bounded derivatives of all
orders
( of smooth bounded functions on  noncommutative  ${\bf R}^d$  ). 
It can be identified 
with  the multiplier algebra of ${\bf R}_{\theta}^d$.(By  definition  the
multiplier 
algebra of $A$ is an endomorphism algebra of $A$ considered as a right 
$A$-module.)

It is important to notice that the above construction can be applied also 
in the case when the topology in the algebra $A$ is specified by a 
countable family of norms.  In particular, we can  take
$A=S({\bf R}^d)$ (algebra of smooth functions 
on ${\bf R}^d$ tending to zero faster that any power); then the
application 
of the above construction gives an algebra denoted by $S({\bf
R}_{\theta}^d)$. 
(Notice, that the norms on $S({\bf R}^d)$ 
are not invariant with respect to translations, 
therefore formally the results of [11] cannot be applied. Nevertheless, 
these results remain correct in this case.)

In the case when $\theta$ is a nondegenerate matrix $d$ is necessarily 
even $(d=2n)$. and we identified $S({\bf R}_{\theta} ^d)$ with the algebra 
of integral operators on ${\bf R}^n$ having kernel belonging to 
$S({\bf R}^d)$.  
 
Various algebras  corresponding to noncommutative ${\bf R}^n$ are 
useful in the analysis of various classes of gauge fields. In particular, 
 gauge fields related to ${\bf R}_{\theta} ^d$ are bounded ,  gauge 
fields corresponding to $\tilde {{\bf R}}_{\theta} ^d$  tend to 
zero at infinity, considering  $S({\bf R}_{\theta} ^d)$ we allow 
 gauge fields that grow polynomially and in the case of 
$\widetilde{S({\bf R}_{\theta }^d)}$ we obtain  gauge fields that 
decrease faster than any power at infinity.

Other interesting versions of noncommutative ${\bf R}^d$ can 
be defined by means of methods of the theory of 
pseudodifferential operators [15]. 
Recall that using the irreducible representation of commutation 
relations $[\hat {x}^k,\hat {x}^l]=i\theta^{kl}$, when $\theta$ 
is a nondegenerate $d\times d=2n\times 2n$ matrix we assign to a 
function $\varphi (x)= \int \varphi(k)e^{ikx}dk$ an operator 
$\hat {\varphi} (x)= \int \varphi(k)e^{ik\hat {x}}dk$ acting on 
functions of $n$ variables. Considering various classes of 
functions $\varphi (x)$ we obtain various classes of 
pseudodifferential operators. For example, we can restrict 
ourselves to the class $\Gamma _{\rho}^m({\bf R}^d)$ of 
smooth functions $a(z)$ on ${\bf R}^d$ obeying 
$$\mid \partial _{\alpha}a(z)\mid \leq C_{\alpha}<z>^{m- 
\rho \mid \alpha \mid }$$ 
where $\alpha =(\alpha _1,...,\alpha_d),\ \  m\in {\bf R}, 
\ \  0<\rho \leq 1, \ \  <z>=(1+\parallel z\parallel ^2)^{1/2}$. 
One can prove that the star-product of functions 
$a^{\prime}\in  \Gamma _{\rho}^{m_1}$ and  
$a^{\prime \prime}\in \Gamma _{\rho}^{m_2}$ belongs to 
 $\Gamma _{\rho}^{m_1+m_2}$.
If $d=2n, \ \  \theta$ is nondegenerate then operators corresponding 
to functions belonging to this class (operators having Weyl 
symbols from $\Gamma_{\rho}^m$) are called 
 pseudodifferential operators of the class $G_{\rho}^m$. 
It follows from the above statement that  the product  $A^{\prime}\cdot
A^{\prime\prime}$
of operators $A^{\prime}\in G_{\rho}^{m_1}$ and
 $A^{\prime\prime}\in G_{\rho}^{m_2}$ belongs to  
$G_{\rho}^{m_1+m_2}$. Using this fact it is easy to 
construct  various algebras  of pseudodifferential operators.

In particular, $\Gamma _{\rho}$ defined as a union of all 
classes $\Gamma_{\rho}^m$ is an algebra with respect to 
star-product; it corresponds to the algebra of 
pseudodifferential operators $G_{\rho}=\cup_m 
G_{\rho}^m$. The class $\Gamma _{\rho}^m$ is an algebra 
for $m\leq 0$.
The intersection $\Gamma _{\rho}^{-\infty}$  of all 
classes $\Gamma _{\rho}^m$ coincides with $S({\bf R}^d)$.

It is important to notice that although algebras 
${\bf R}_{\theta} ^d$, $S({\bf R}_{\theta} ^d)$,  
$\Gamma _{\rho}^m$, $m<0$,
 are different and are related to  
different classes of gauge fields they have the same 
$K$-theory. For non-degenerate $\theta $ these algebras 
are closely related to the algebra $K$ of compact 
operators on Hilbert space. One can show that not 
only elements of $K$-theory groups, but also 
projective modules over unitized algebras 
$\tilde {{\bf R}}_{\theta} ^d$, $\widetilde 
{S({\bf R}_{\theta} ^d})$, $\tilde {\Gamma} _{\rho}^m$, 
$m<0$, are in one-to-one correspondence with projective 
modules over $\tilde {K}$. This is the reason why it 
is possible to use $K$ and $\tilde {K}$ in the 
consideration of noncommutative solitons and instantons. 
If $K$ is defined as an algebra of compact operators 
acting  on Hilbert space ${\cal H}$, then the module 
$K^1$ can be considered as (completed) 
direct sum of countable numbers of copies of ${\cal H}$ 
regarded as a $K$-module. It follows from this 
representation that ${\cal H}^k\oplus K^1$ 
is isomorphic to $K^1$ (if we add $k$ elements to a 
countable set we again obtain a countable set) and that 
$K^n$ is isomorphic to $K^1$. Every projective module 
over $\tilde {K}$ is isomorphic to a direct sum of 
several copies of ${\cal H}$ and several copies of 
free module $\tilde {K}^1$ [see Appendix]. We used analogs 
of these statements in Sec.2 and 3.

  Let us say that a function $a(z)$ belongs to the class  
$\tilde {\Gamma} _{\rho}^m({\bf R}^d)$ if $\mid a(z)\mid 
\leq{\rm const}\cdot <z>^m,\ \   \mid \partial _{\alpha}a(z)\mid  
<C_{\alpha}\mid a(z)\mid <z>^{-\rho \mid \alpha \mid }$. It is easy to
check that 
$\tilde {\Gamma} _{\rho}^m \subset  \Gamma _{\rho}^m$. We'll 
say that $a\in H  \Gamma _{\rho}^{m,m_0}$  if  $a \in 
\tilde {\Gamma }_{\rho}^m$  and  there exists such a function 
$b\in \tilde {\Gamma} _{\rho}^{-m_0}$ that $a\cdot b=1-r, 
\ \  r\in S({\bf R}^d)$.  (In other words, the inverse to the function 
$a(z)$ with respect to usual multiplication should exist for large 
$\mid z \mid$ and belong to  $\tilde {\Gamma} _{\rho}^{-m_0}$.)  
 One can prove that for a  function  $a\in H \Gamma _{\rho}^{m,m_0}$ 
there exists a function   $b_{\theta }\in  H \Gamma _{\rho}^{-m,-m_0}$ 
(parametrix)  obeying $1-a\star b_{\theta}\in S({\bf R}^d), \ \  
1-b_{\theta}\star a \in S({\bf R}^d)$. Here the star -product depends 
on the noncommutativity parameter $\theta$. (One can say that a 
function  $a(z)$ that is invertible up to an element of $S({\bf R}^d)$ 
with respect to the usual multiplication has this property also for the 
star-product.)

Notice that the parametrix is essentially unique: if
$b_{\theta}^{\prime}\in 
 \Gamma_{\rho}$  and $1-a\star  b_{\theta}^{\prime}\in S({\bf R}^d)$
then  $b_{\theta}-b_{\theta}^{\prime}\in S({\bf R}^d)$ (and therefore   
 $1-b_{\theta}^{\prime} \star a\in S({\bf R}^d)$).

Let us assume that  pseudodifferential operator $\hat {a}$ corresponding 
to a function $a\in H\Gamma_{\rho}^{m, m_0}$ has no zero modes 
(${\rm Ker} \hat {a}=0$). Then we can construct an operator 
$\hat {T}=\hat {a}(\hat {a}^+\hat {a})^{-1/2}$ 
obeying $\hat {T}^+\hat {T}=1$. It is easy to check that corresponding
functions (symbols) $T$ and $T^+$ belong to 
$H\Gamma_{\rho}^{0,0}$. It follows from uniqueness of  parametrix that 
$T^+$  is a  parametrix  of $T$ , i. e. $1-TT^+$ is an integral operator
with a 
kernel from Schwartz space. This remark gives us a way to construct an 
endomorphism  $T$ entering the definition of a field that is gauge trivial 
at infinity.  

  The definition of classes $\Gamma _{\rho}^m,\ \   \tilde {\Gamma
}_{\rho}^m, 
\ \  H \Gamma _{\rho}^{m,m_0}$ was formulated in such a way that it makes 
sense also for matrix valued functions. (The absolute value should be 
understood as a matrix norm.) Such functions can be considered as matrices 
with entries from $\Gamma _{\rho}^m$;  multiplication  of such matrices 
is defined in terms of star-product and  usual matrix  multiplication.
All
statements listed above (including the existence of parametrix) remain 
true also in the matrix case.   

An $s\times s$ matrix $A$ belonging to $H \Gamma _{\rho}^{m,m_0}$
can be considered as an operator acting on $( \Gamma _{\rho})^s $ where   
$ \Gamma _{\rho}=\cup_m  \Gamma _{\rho}^m$. 
 One can prove that ${\rm Ker} A\subset  (S({\bf R}_{\theta}^d))^s$. If 
$\theta $ is a nondegenerate matrix then ${\rm Ker }A$, considered as 
$\tilde {{\bf R}}_{\theta}^d$-module, is isomorphic to ${\cal F}^k$.  
(If  $A\varphi =0$, or, more generally, $A\varphi\in (S({\bf R}^d))^s$ 
it follows from the existence of parametrix that $\varphi \in  
(S({\bf R}^d))^s $. To check that 
${\rm Ker} A$ is isomorphic to ${\cal F}^k$  it is convenient to 
consider $A$ as a pseudodifferential operator $\hat {A}$; using the 
parametrix one can derive that  $\hat {A}$ is a Fredholm operator. 
The relation  $A\varphi =0$  implies that  $\hat {A} \hat {\varphi} =0$; 
in other words $\hat {\varphi}$   is an operator taking values in 
finite-dimensional vector space  ${\rm Ker }\hat {A}$. This 
description  leads to identification  of ${\rm Ker}A$
with ${\cal F}^k$ where $k={\rm dim Ker}\hat {A}$.) 

Let us  consider an operator ${\cal  A}$ acting from  $( \Gamma
_{\rho})^{s+t} $
into  $( \Gamma _{\rho})^s $  and transforming a pair $(u,v)$ where $u\in 
  ( \Gamma _{\rho})^s ,\ \  v\in ( \Gamma _{\rho})^t $ into $Au+Cv\in  
 ( \Gamma _{\rho})^s $.  Let us suppose that the operator $A$ is
represented  
by an $s\times s$ matrix,  belonging to $H \Gamma _{\rho}^{m,m_0}$, where 
$m>0$, and $C$ is represented by an $s\times t$ matrix belonging to  
$( \Gamma _{\rho})^0$. We denote by $Q$ a parametrix of $A$ (i.e. 
$QA=1-\Pi ,\ \  AQ=1-\Pi ^{\prime}$, and1 $\Pi, \Pi ^{\prime}\in 
S({\bf R}^d)$). 

  We would like to construct an  ${\bf R}_{\theta} ^d$-linear isomorphism 
between  ${\bf R}_{\theta} ^d$-modules ${\cal E}={\rm Ker}{\cal A} \cap  
 ({\bf R}_{\theta} ^d)^{s+t}$  and ${\cal E}^{\prime}={\cal F}^k\oplus 
({\bf R}_{\theta} ^d)^t$. Let us assume first that $QA=1$ (i.e. 
$\Pi^{\prime}=0$). Then $Au+Cv=0 $ implies $u=u_0-QCv$ where 
$u_0\in {\rm Ker} A={\cal F}^k$. Let us check that the map $\nu$
transforming $(\rho ,v)$ into $(\rho-QCv,v)$ is  an  isomorphism 
between ${\cal E}^{\prime}$ and ${\cal E}$. This map is obviously  
 ${\bf R}_{\theta} ^d$-linear and has trivial kernel. To check that 
the image of a point $(\rho ,v)\in {\cal E}^{\prime}$ belongs to 
${\cal E}$ we notice that $Q\in H\Gamma_{\rho}^{-m, -m_0}\subset 
\Gamma_{\rho}^{-m}$and therefore $QC\in \Gamma_{\rho}^{-m}$. 
We assumed that $m>0$, hence $-m<0$ and $v\in  
({\bf R}_{\theta} ^d)^s$ implies $QCv\in  ({\bf R}_{\theta} ^d)^s$. 
Conversely, the map $\nu ^{-1}: (u,v)\rightarrow (u+QCv,v)$ transforms 
${\cal E}$ into ${\cal E}^{\prime}$ . 

 The above consideration clarifies and generalizes the arguments 
applied in Sec.3 to the case when $A\in  H\Gamma_1^{1,1}$ and 
$C\in \Gamma _1 ^0$ are specified by the formula (23). 

Now we should get rid of the condition $\Pi ^{\prime}=0$. This can be 
done by means of change of variables: $\tilde {u}=u, \ \  \tilde
{v}=v+Mu$, 
where operator $M$ is selected in such a way that the above arguments 
can be applied to the equation $(A-CM)\tilde {u}+C\tilde {v}=0$.

If a commutative Lie algebra ${\cal L}={\bf R}^d$ acts on associative 
algebra $A$ by means of derivations we can define a connection on a 
(right) $A$-module $E$ as a collection of ${\bf C}$-linear operators 
$\nabla _1,...,\nabla _d$ acting on $E$ and obeying the Leibniz rule:
 \begin {equation}
\nabla _i(ea)=\nabla _ie\cdot a+e\delta _ia
\end {equation}
where $e\in E,\  a \in A,\  \delta _i:A\rightarrow A$ are derivations 
(infinitesimal automorphisms) corresponding to elements of a basis of 
${\cal L}$. A curvature of a connection (field strength of noncommutative 
gauge field) is defined by the formula
 \begin {equation}
F_{ij}=[\nabla_i,\nabla _j]
\end {equation}
  It is easy to check that $F_{ij}:A\rightarrow A$ are linear maps 
(endomorphisms of $A$-module $E$). One can consider the curvature as a 
$2$-form $F$ on ${\cal L}={\bf R}^d$ taking values in the algebra of 
endomorphisms ${\rm End}_AE$. 

We'll work with involutive algebras and Hilbert modules (i.e. we assume 
that  the algebra $A$ is equipped with antilinear involution $a\rightarrow 
a^*$ and the modules are provided with $A$-valued inner product). Then it 
is natural to consider gauge fields as unitary connections; the 
endomorphisms $F_{ij}$ 
will be antihermitian operators in this case. 

If this algebra is equipped with a trace and ${\cal L}={\bf R}^d$ is 
equipped with a non-degenerate inner product (with metric tensor 
$g_{ij}$) then we can define Yang-Mills action functional on 
connection on the $A$-module $E$ 
 \begin {equation}
S={\rm Tr}F_{ij}F^{ij}={\rm Tr}<F,F>.
\end {equation}
(The indices are raised and lowered by means of metric tensor $g_{ij}$.) 

In particular, we can apply the construction of action functional to the 
cases when $A=T_{\theta}^d,{\bf R}_{\theta}^d$ or $\tilde 
{{\bf R}}_{\theta}^d$.

Using the  metric tensor $g_{ij}$ on ${\cal L}$ we can define the Hodge 
dual of an exterior form on ${\cal L}$. In the case $d=4$ we can define 
an instanton (antiinstanton) as a gauge field obeying 
\begin {equation}
F\pm *F=\omega\cdot 1
\end {equation}
where $\omega$ is a ${\bf C}$-valued $2$-form on ${\cal L}$. In other 
words, the self-dual part of the curvature of an instanton should be a 
scalar (or, more precisely, a scalar multiple of unit endomorphism). 
In the case of an antiinstanton this condition should be fullfilled 
for the antiselfdual part of the curvature. We suppose that the metric 
on ${\cal L}$ used in the definition of (anti)instanton is positive. 

Let us introduce a complex structure on ${\cal L}={\bf R}^4$ in such 
a way that ${\cal L}$ becomes a Kaehler manifold. Without loss of 
generality we can assume the metric has the form $ds^2=dz_1d 
\bar {z}_1+dz_2d\bar {z}_2$ in complex coordinates $z_1,z_2$ on 
${\cal L}$. Then instead of operators $\nabla_1,\  \nabla_2,\  
\nabla_3,\  \nabla_4$ corresponding to the coordinates $x_1,\  
x_2, x_3,\  x_4$ on ${\cal L}$ it is convenient to consider 
operators $D_1=\nabla_1+i\nabla _2, D_2=\nabla_3+i\nabla _4, 
D_1^*=-\nabla_1+i\nabla _2, D_2^*=-\nabla_3+i\nabla _4$. (Here 
$z_1=x_1+ix_2,\  z_2=x_3+ix_4$.) The instanton equation 
$F+*F=\omega\cdot 1$ can be represented in components 

$$F_{11}+F_{34}=\omega _{11}\cdot 1$$

$$F_{14}+F_{23}=\omega _{14}\cdot 1$$

$$F_{13}+F_{42}=\omega _{13}\cdot 1$$

or, equivalently, in the form 
\begin {equation}
[D_1,D_2]=\lambda\cdot 1
\end {equation}
\begin {equation}
[D_1,D_1^+]+[D_2,D_2^+]=\mu\cdot 1.
\end {equation}

In noncommutative case one can describe the moduli space 
of instantons in terms of holomorphic vector bundles 
(Donaldson's theorem). There exists an analog of this 
description for noncommutative instantons; the role of 
holomorphic bundles is played by solutions of (31).

Let us consider the case when $A=T_{\theta}^4$ and $E$ is a 
projective module (a direct summand in a free module). There 
exists a natural trace on $T_{\theta}^d$ and this trace 
induces a trace on the algebra of endomorphisms ${\rm End}_AE$. 
(Endomorphisms of a free module over a unital algebra $A$ can 
be represented as matrices with entries from $A$. Trace of 
such a matrix can be defined as a sum of traces of 
diagonal elements. An endomorphism of a projective module can 
be extended to an endomorphism of a free module. This permits 
us to define a trace of an endomorphism of projective module.)

 It is easy to check that ${\rm Tr}F_{ab}$ does not depend on the 
choice of connection $\nabla$ on $T_{\theta}^d$-module $E$. To 
derive this statement we can use the formula for infinitesmal 
variation of $F_{ab}$:
$$\delta F_{ab}=[\nabla _a,\delta \nabla _b]+[\delta\nabla _a, 
\nabla _b].$$ It remains to use the fact that 
\begin {equation}
{\rm Tr}[\nabla _a,\varphi]=0
\end {equation}
for every endomorphism $\varphi$. (By definition, the trace vanishes 
on commutator of endomorphisms. The difference $\nabla _a^{\prime}-
\nabla _a$ where $\nabla^{\prime},\  \nabla$ are two connections is an 
endomorphism, hence it is sufficient to check (32) only for one 
connection. For example, one can check (32) for so called Levi-Civita 
connection $P\delta _iP$ where $P$ is a projection of free module 
on the projective module $E$.)  We see that ${\rm Tr}F_{ab}$ 
is a topological number characterizing the $T_{\theta}^d$-module 
$E$. Other numbers of this kind can be obtained by the 
formula ${\rm Tr}F^n$ where $F$ is considered as a $2$-form on 
${\cal L}$. To prove that ${\rm Tr}F^n$  does not depend on the 
choice of connection we again should use (33).  If $d=4$ we have the 
following topological numbers: ${\rm Tr}1$  (noncommutative 
dimension), ${\rm Tr}F_{ab}$ (magnetic flux) and ${\rm Tr}F^2$. 
For the case of commutative torus  $(2\pi)^{-n}{\rm Tr}F^n$ 
is an integer. 
In the case of noncommutative torus $T_{\theta}^d$ also one can 
find integer  numbers characterizing topological class (more 
precisely, $K$-theory class) of the module $E$: as in commutative 
case we have integer-valued antisymmetric tensors of even rank: 
$\mu^{(0)}, \mu_{ij}^{(2)}, \mu_{ijkl}^{(4)},... $. ($K$-theory 
group is a discrete object; it does not change by continuous variation  
of   $\theta$). These numbers can be related to ${\rm Tr} F^n$.

One can prove that the value of action functional (29)  on a gauge field
obeying (30) (on an (anti)instanton) can be expressed in terms of 
topological numbers of the module $E$  and that this value gives a 
minimum to the action functional on gauge fields on $E$. The proof 
is based  on the inequality
\begin {equation}
{\rm Tr}(F\pm\ast F-\omega\cdot 1)(F\pm\ast F-\omega 
\cdot 1)^{\ast}\geq 0
\end {equation}
that follows from the positivity of trace (from relation ${\rm Tr} 
A^{\ast}A\geq 0$). The left hand side of (34) can be expressed in 
terms of the value of action functional and of topological numbers 
of the module $E$; this gives the estimate we need (see [7] for detail) 
The above consideration looks very general; however, we encounter 
some problems trying to apply it to ${\bf R}_{\theta}^4$  or to 
$\tilde {{\bf R}}_{\theta}^4$. First of all, one can define a trace 
of element  $f\in {\bf R}_{\theta}^4$ as an integral of $f$ over 
${\bf R}^4$, but this integral can be divergent. This trace is positive; 
we can extend it to a trace on $\tilde {{\bf R}}_{\theta}^4$  
assigning an arbitrary value to   ${\rm Tr}1$, but the extended 
trace is not positive. Therefore in Sec 1 and 2 we defined instantons 
with $\omega =0$ in  the right hand side of (30). However, the 
topological number ${\rm Tr}F^2$  is well defined for instantons on 
 ${\bf R}_{\theta}^4$. (This follows from finiteness of euclidean 
action.) The above arguments show that the euclidean action of an 
instanton on  ${\bf R}_{\theta}^4$ is minimal in the set of gauge 
fields with given topological number ${\rm Tr}F^2$ . It is important 
to notice that the topological number ${\rm Tr}F$  is ill-defined in 
this situation (the corresponding integral diverges).

      {\bf  Acknowledgements.} I am deeply indebted to  A.
Konechny, N.Nekrasov, and M.Shubin for very useful 
discussions. My special thanks to A.Connes and M.Rieffel 
for their help. 

   \centerline {{\bf Appendix}}

\centerline {{\bf Projective modules over  $\tilde {K}$}}

Let us consider the $C^\ast$-algebra $K$ of compact operators
on Hilbert space ${\cal H}$ and corresponding unitized
algebra $\tilde {K}$.  One can prove the following statement.

{\it Every finitely generated projective module
over $\tilde {K}$ is isomorphic to a direct sum of
several copies of ${\cal H}$ and several copies of
free module $\tilde {K}^1$.}

Proof. Let $\rho$ be the canonical character of  $\tilde {K}$
(i.e. the canonical map $\tilde {K} \rightarrow \tilde {K}/ K=\bf C $).
Let $e \in M_q (\tilde {K})$ be a selfadjoint idempotent and
$\rho(e) \in M_q (\bf C)$ the corresponding scalar idempotent.
The density of finite rank operators in $K$ shows that for any
$\epsilon >0$ one can find a closed subspace $E$ of finite codimension
 in ${\cal H}$ and a selfadjoint $x \in M_q (\tilde {K})$ such that the
norm of $x - e$ is less than $\epsilon $ and the matrix elements of
$ x - \rho(e)$ all vanish on $E$. Viewing $x$ as an element of
$M_q (\tilde {A})$ where A is the matrix algebra of operators in the
orthogonal
of $E$ one gets that for $\epsilon >0$ small enough, $x$ is close to a
self-adjoint idempotent $f \in M_q (\tilde {A})$ and that $f$
is equivalent to $e$ in $M_q (\tilde {K})$. The result then follows
from the trivial determination of the finite projective modules
on $\tilde {A}$.

\vskip .25in
\centerline {\bf References}
\vskip .1in
    1. A. Connes, {\it  C*-algebres et Geometrie Differentielle}, C. R.
Acad. Sci. Paris {\bf 290} (1980), 599-604, English translation
hep-th/0101093 .

    2. A. Connes,  M. Douglas,  and  A. Schwarz,  {\it  Noncommutative
Geometry and Matrix Theory: Compactification on  Tori},  JHEP {\bf 02}
(1998),  3-38

    3. A. Konechny, A. Schwarz, {\it Introduction to M(atrix) Theory and 
Noncommutative Geometry},   hep-th/0012145
   
   4. A. Schwarz, {\it Gauge Theories on Noncommutative Spaces}, Lecture
on 
ICMP -2000,  hep-th/0011261

   5. N. Seiberg and E. Witten, {\it String Theory and Noncommutative
Geometry}, JHEP {\bf 09} (1999),  032.   

   6. N. Nekrasov  and  A. Schwarz,  {\it Instantons on Noncommutative
$R^4$ and (2,0) Superconformal Six-dimensional Theory}, Comm. Math.
 Phys. {\bf 198} (1998), 689-703.

   7.  A. Astashkevich,  N. Nekrasov   and  A. Schwarz,  {\it On
Noncommutative Nahm Transform}, Comm. Math. Phys. {\bf  211 (1)} (2000),
167-182.

   8. D. Gross, N. Nekrasov, {\it Monopoles and 
strings in noncommutative gauge theory}, hep-th/0005204  {\it Dynamic 
of strings in noncommutative gauge theory}, hep-th/0007204  {\it 
Solitons in noncommutative gauge theories}, hep-th/0010090

   9. N. Nekrasov, {\it Noncommutative Instantons Revisited},
hep-th/0010017 {\it Trieste Lectures on Solitons in Noncommutative
Gauge Theories}, hep-th/0011095

   10. A. Connes, {\it Noncommutative Geometry},  Academic Press (1994)
1-655

    11. M. Rieffel,  {\it Deformation Quantization for Actions of ${\bf
R}^d$}, 
Memoirs of AMS, {\bf 106 (506)}  (1993),1-93

     12 C. I. Lazaroiu. {\it    A noncommutative-geometric interpretation of the resolution of
equivariant instanton moduli spaces},
 hep-th/9805132

     13. J. A. Harvey, P. Kraus, and  F. Larsen {\it Exact Noncommutative
Solutions}, 
hep-th/0010060

      14.  A. Konechny, A. Schwarz, in preparation.

      15.  M. Shubin, {\it Pseudodifferential  operators}

      16. K. Furuuchi, {\it Instantons on Noncommutative ${\bf R}^4$ 
and Projection Operators}, Prog. Theor. Phys. {\bf 103} (2000) 1043,
hep-th/9912047  {\it Equivalence of Projections as Gauge Equivalence  
on Noncommutative Space}, hep-th/0005199  {\it Topological Charge of
$U(1)$ Instantons}, hep-th/0010006

      17.M. Aganagic, R. Gopakumar, S. Minwalla, A. Strominger
    {\it Unstable Solitons in Noncommutative Gauge Theory}
 hep-th/0009142

\end {document}